\documentclass[letterpaper,10pt,journal,twoside]{IEEEtran}

\usepackage{graphicx}
\usepackage{booktabs}
\usepackage{amsmath}
\usepackage{url}
\usepackage[hidelinks]{hyperref}

\title{Fly Whiskers Meet LiDAR: Robust Navigation\\ under Sensor Failure}

\author{Ali Mahzarnia and Homayoun Jamshidi%
\thanks{A. Mahzarnia and H. Jamshidi are independent researchers (e-mail: mhznariman.ali@gmail.com; homayoun.jamshidi@gmail.com). Both authors contributed equally to this work.}%
}

\begin{document}

\maketitle

\begin{abstract}
We extend FLYNN, a leaky integrator recurrent network wired on the Drosophila melanogaster connectome, with a 32-ray 2D LiDAR interface that requires no learned fusion layer beyond a scalar input gain. A 270 degree frontal scan at 8 m range is injected as inverted proximity into the 37 BM\_Vib head-bristle neurons. Across 24 paired comparisons in a four-state vision matrix with LiDAR on or off, FLYNN-LiDAR has 17 significant wins, 5 nonsignificant comparisons, and 2 significant losses against three LiDAR-capable baselines. Both losses are to MobileNet with the left eye off. With full vision, FLYNN-LiDAR reaches 96.6\% success. Under total vision loss, the scan raises success from 45.2\% to 64.6\%; the baselines reach 44.6\% to 56.6\% with the scan on. Under out of distribution textures, it is the only LiDAR-capable model whose success improves with the scan (44.2\% to 58.4\%); both CNNs score 0\%, and SmallWorld declines from 26.8\% to 7.4\%. Under periodic vision dropout, FLYNN-LiDAR achieves 93.6\% and 89.6\% success at 25\% and 50\% duty with fewer collisions. Profiling the evaluated forward paths gives 3.93M multiply accumulate operations per step; EfficientNet and MobileNet require 4.60$\times$ and 2.94$\times$ as many. Final-state analysis shows near-linear sensory superposition (cosine 0.9955).
\end{abstract}


\section{Introduction}

\IEEEPARstart{F}{LYNN}~\cite{flynn2026} turns the adult Drosophila connectome into a leaky integrator recurrent network that navigates a simulated robot around obstacles. Its visual and tactile inputs occupy biologically identified pathways, and its descending-neuron readout is trained by DAgger~\cite{ross2011dagger}. We ask whether a new range modality can enter through an existing nonvisual pathway and improve navigation when vision fails.

Head bristles are the dipteran whisker analog. Their vibration-sensitive BM\_Vib afferents report near-field mechanosensory input, but receive no direct sensory drive in FLYNN. We inject a fixed, spatially ordered projection of a 32-ray 2D LiDAR scan into these 37 neurons. The projection has fixed weights and one learned scalar input gain; integration occurs in the recurrent network.

The strongest benefits occur under total vision loss, out of distribution (OOD) textures, and periodic dropout. The result is not universal superiority: MobileNet retains a significant left-eye-off advantage, and the two models show opposite left/right asymmetries. We therefore report the full matrix and paired tests rather than a single aggregate ranking. A browser demo is publicly available at \url{https://ali-mahzarnia.github.io/flynn-lidar/}.

Our contributions are:
\begin{itemize}
\item A fixed 32-ray to 37-neuron projection through the fly BM\_Vib whisker pathway, with true anatomical laterality and only a learned scalar input gain.
\item A paired evaluation with 500 seeds per matrix and OOD cell and 250 per blackout or flicker cell, including explicit analysis of lateral asymmetry and nonsignificant cells.
\item An operation count based on the evaluated coalesced sparse graph and the actual $30\times60$ CNN input, with per-episode energy estimates from saved durations.
\item Correlational final-state analyses of sensory superposition and LiDAR-driven state displacement on 100 seeds per condition.
\end{itemize}

\section{Related Work}

Connectome-constrained models support visual-response prediction~\cite{lappalainen2024connectome}, whole-brain sensorimotor simulation~\cite{shiu2024computational}, and robot control~\cite{flynn2026}. Sensor-failure robustness has been studied with learned gating~\cite{patel2019gated} and out of distribution liquid networks~\cite{chahine2023liquid}. We instead add range input through a biological whisker channel and compare against both CNNs and a Watts-Strogatz small-world control~\cite{watts1998smallworld}.

\section{Methods}

\subsection{Base model}

FLYNN uses FlyWire neurons and recurrent weights in a leaky integrator RNN~\cite{flynn2026,dorkenwald2024flywire,schlegel2024wholebrain}. Camera pixels enter through a virtual retina at photoreceptor-column positions, tactile bristles report contact, goal direction enters through Johnston's organ neurons, and descending neurons drive a small velocity and heading readout. We retain the original model equations; model-specific training configurations are detailed in Section~\ref{sec:repro}.

\subsection{LiDAR interface}

The sensor is a 32-ray 2D scan spanning a 270 degree frontal fan at 8 m maximum range, updated once per 50 Hz control step. It represents a downsampled RPLIDAR-class spinning scanner~\cite{slamtec_rplidar_a1}. Each continuous range is clipped to $[0,8]$ m and encoded as $1-r/8$, so zero denotes free space and one denotes contact. Rays are assigned by hemifield to the nearest of 18 left and 19 right BM\_Vib afferents in a fixed midline-to-lateral map. The projection has no learned weights; the network has a scalar input gain for the modality. BM\_Vib does not overlap the separate tactile head-bristle input.

\begin{figure}[!t]
\centering
\includegraphics[width=\columnwidth]{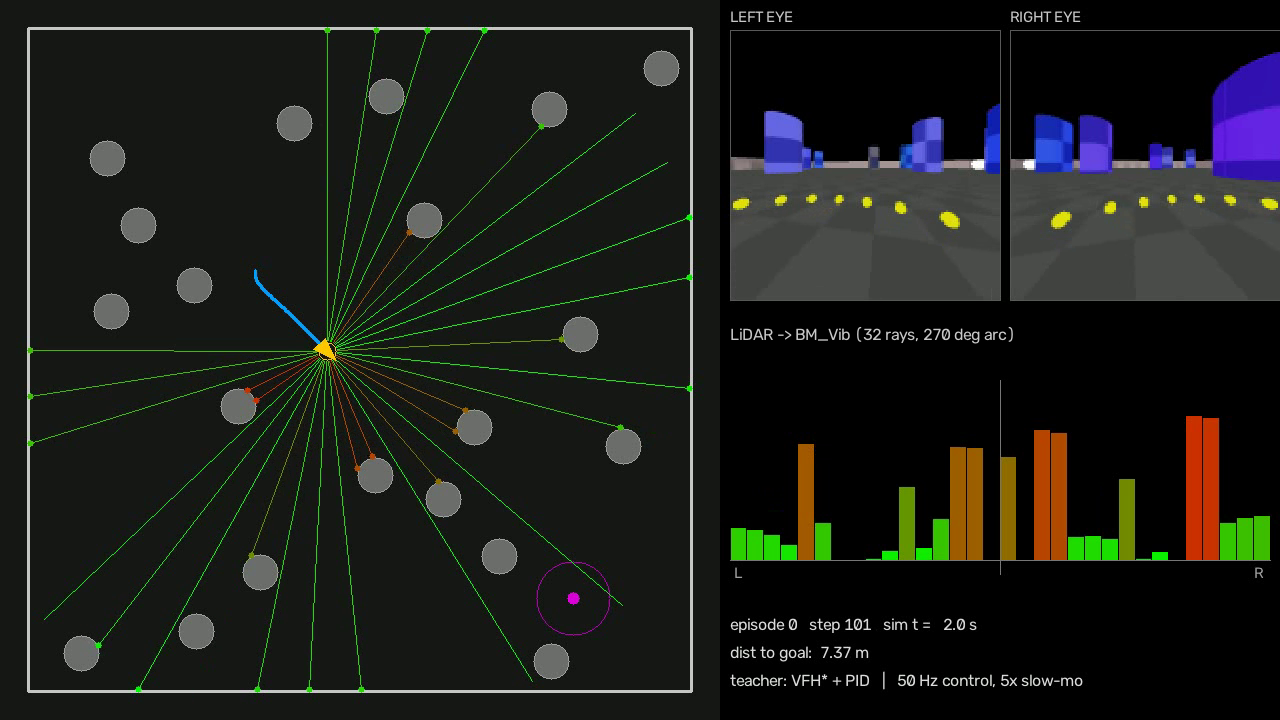}
\caption{Sensory suite: arena view with the 32-ray 270 degree LiDAR fan, left and right camera frames, and the inverted scan. The fixed projection maps the rays onto 37 BM\_Vib afferents (18 left, 19 right).}
\label{fig:suite}
\end{figure}

In the connectome model, the LiDAR-to-whisker mapping preserves true anatomical laterality; in the small-world control, laterality is assigned by convention, as the graph carries no spatial anatomy. The control receives 37 arbitrarily labeled sites with the same 18/19 split. Each CNN receives the identical 32-dimensional vector as additional input. Fig.~\ref{fig:suite} shows the sensory suite.

\subsection{Training}

All policies use four DAgger iterations against an analytic VFH* planner with PID tracking~\cite{ross2011dagger,ulrich2000vfh}. Teacher mixing and action noise decay to zero over the schedule. For student command $(\hat v,\hat\theta)$ and teacher command $(v,\theta)$, each post-burn-in sample contributes
\begin{equation}
\ell=(\hat v-v)^2+2\left[1-\cos(\hat\theta-\theta)\right],
\end{equation}
and training minimizes the sample mean. The first term is velocity MSE; the heading term costs 0 for no angular error, 2 at 90 degrees, and 4 at 180 degrees.
EfficientNet and MobileNet receive camera and LiDAR dropout augmentation. The connectome and SmallWorld models receive no sensor dropout. Realized episode and optimizer budgets differ from the nominal schedule for every model and are listed in Section~\ref{sec:repro}.

\subsection{Evaluation protocol}

Experiments use MuJoCo~\cite{todorov2012mujoco}, a 14 by 14 m arena, 20 random obstacles, and a random goal. Success is final goal distance below 0.8 m. Episodes have at most 600 steps at $dt=0.02$ s. The matrix crosses full vision, left eye off, right eye off, and total loss with LiDAR on or off. Disabled cameras are zeroed. Each cell uses the same 500 seeds across models: 0 to 249 and 1000 to 1498 with stride 2. OOD evaluation replaces the checkerboard with photorealistic textures. Blackout starts total vision loss at 0, 2, 4, or 6 s with LiDAR on. Flicker drops both eyes in a 4 s square wave at 25\% or 50\% duty; these sweeps use seeds 0 to 249.

We report Wilson 95\% intervals and continuity-corrected paired McNemar tests. The 24 matrix tests compare FLYNN-LiDAR with three LiDAR-capable baselines over eight cells and use a Bonferroni correction. The 12 blackout baseline-by-onset tests form a second corrected family. Nonsignificance is not treated as evidence of equivalence.

\subsection{Implementation details}
\label{sec:impl}

The recurrence is $h_{t+1}=a h_t+(1-a)\tanh(W h_t+x_t+b)$, with learned retention $a$ per cell class~\cite{flynn2026}. In the evaluated FLYNN checkpoints, 5{,}342{,}446 connection records coalesce to 3{,}732{,}460 nonzero recurrent weights. Including these weights, 138{,}584 biases, 8{,}542 cell-class leak logits, readout and wind MLPs, and scalar gains gives 4.08M learned parameters for FLYNN-LiDAR and FLYNN (repro.). SmallWorld has 138{,}603 nodes, 5{,}405{,}517 recurrent weights, 1{,}303 rather than 1{,}288 readout inputs, and 5.76M parameters. MobileNet and EfficientNet have 3.99M and 5.29M parameters, respectively.

Each CNN resizes the eyes to $30\times30$ grayscale, concatenates them into one $30\times60$ image, and uses one backbone: 960 features for MobileNetV3~\cite{howard2019mobilenetv3} or 1280 for EfficientNet-B0~\cite{tan2019efficientnet}. Features, scan, two-dimensional goal direction, and collision flag enter a 256-unit GRU, giving 995 or 1315 inputs. The nominal schedule is 500 episodes and 300 optimizer steps per iteration at batch 128, with balanced sampling over straight, turn, collision, pre-collision, and start categories.

\section{Experiments and Results}

\subsection{Robustness matrix}

\begin{figure*}[!t]
\centering
\includegraphics[width=\textwidth]{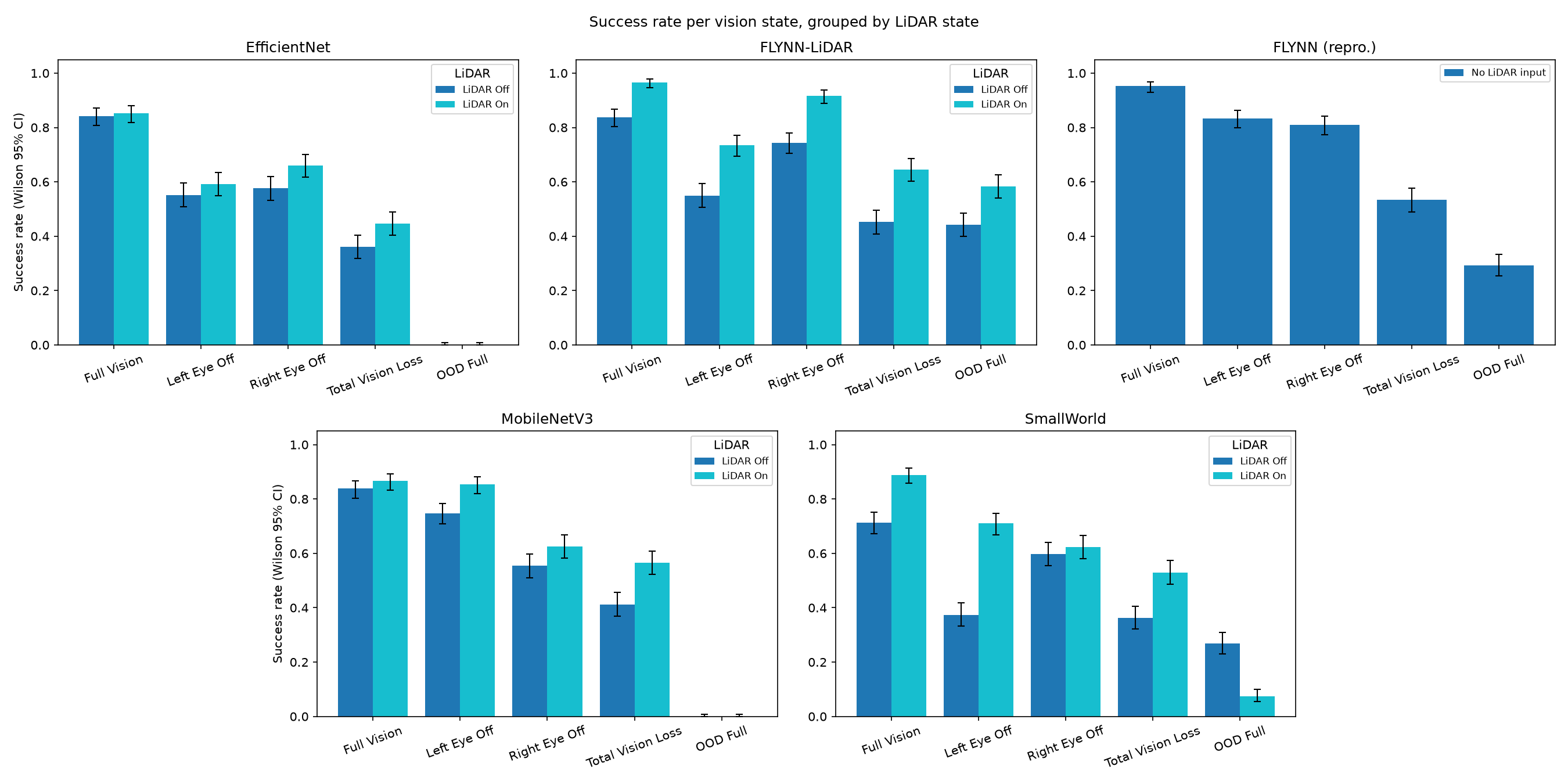}
\caption{Success rates for four vision states and OOD textures with LiDAR on or off, 500 paired seeds per cell. Error bars are Wilson 95\% intervals. FLYNN (repro.) has no LiDAR input, so one bar per state is shown. FLYNN-LiDAR does not lead in every cell: MobileNet has the highest left-off success among the LiDAR-capable models.}
\label{fig:matrix}
\end{figure*}

\begin{table}[!t]
\caption{Success rate (\%) and mean collisions per episode, LiDAR on / off, 500 paired seeds per cell. Bold marks the highest success in each LiDAR-state column. FLYNN (repro.) has no LiDAR input, so a single value is shown per cell. Wilson 95\% intervals are within $\pm$4.4 points. Collision counts must be read with success because stalled robots can avoid contact.}
\label{tab:matrix}
\centering
\scriptsize
\setlength{\tabcolsep}{3.5pt}
\begin{tabular}{lccccc}
\toprule
 & Full & Left off & Right off & Total loss & OOD \\
\midrule
\multicolumn{6}{l}{\emph{Success rate (\%)}} \\
\textbf{FLYNN-LiDAR} & \textbf{96.6} / 83.8 & 73.4 / 55.0 & \textbf{91.6} / 74.4 & \textbf{64.6} / 45.2 & \textbf{58.4} / \textbf{44.2} \\
FLYNN (repro.) & 95.2 & 83.4 & 81.0 & 53.4 & 29.2 \\
EfficientNet & 85.2 / 84.2 & 59.2 / 55.2 & 66.0 / 57.6 & 44.6 / 36.0 & 0.0 / 0.0 \\
MobileNet & 86.6 / 83.8 & \textbf{85.4} / 74.8 & 62.6 / 55.4 & 56.6 / 41.2 & 0.0 / 0.0 \\
SmallWorld & 88.8 / 71.4 & 71.0 / 37.4 & 62.4 / 59.8 & 53.0 / 36.2 & 7.4 / 26.8 \\
\midrule
\multicolumn{6}{l}{\emph{Mean collisions per episode}} \\
\textbf{FLYNN-LiDAR} & 2.3 / 19.3 & 26.7 / 67.6 & 4.9 / 32.3 & 40.1 / 80.5 & 11.4 / 36.4 \\
FLYNN (repro.) & 2.0 & 4.8 & 6.7 & 13.9 & 49.4 \\
EfficientNet & 8.3 / 8.6 & 58.6 / 63.4 & 55.6 / 67.3 & 47.4 / 67.7 & 4.9 / 9.7 \\
MobileNet & 7.7 / 8.6 & 19.8 / 36.6 & 66.2 / 58.6 & 65.3 / 219.9 & 5.3 / 16.3 \\
SmallWorld & 5.0 / 35.6 & 28.5 / 93.2 & 45.6 / 74.1 & 51.5 / 106.3 & 4.7 / 44.9 \\
\bottomrule
\end{tabular}
\end{table}

Table~\ref{tab:matrix} and Fig.~\ref{fig:matrix} give the complete comparison. Against EfficientNet, MobileNet, and SmallWorld, FLYNN-LiDAR has 17 significant wins, 5 nonsignificant comparisons, and 2 significant losses after Bonferroni correction across 24 tests. Both losses are to MobileNet with the left eye off: 73.4\% versus 85.4\% with LiDAR and 55.0\% versus 74.8\% without. The nonsignificant cells are full-vision LiDAR-off against either CNN, left-off LiDAR-off against EfficientNet, left-off LiDAR-on against SmallWorld ($p=0.331$), and total-loss LiDAR-off against MobileNet ($p=0.106$).

With LiDAR on, FLYNN-LiDAR leads the best of the three LiDAR-capable baselines by 7.8 points at full vision and by 8.0 to 20.0 points under total loss. Its 1.4-point full-vision lead over FLYNN (repro.) is not significant ($p=0.296$). All three total-loss LiDAR-on comparisons remain significant after correction (largest adjusted $p=0.0323$). Its within-model LiDAR gains are 12.8, 18.4, 17.2, 19.4, and 14.2 points at full, left-off, right-off, total loss, and OOD, respectively; each has five-test Bonferroni-adjusted $p<10^{-6}$. FLYNN-LiDAR was trained with LiDAR always on, so its weak LiDAR-off cells are an ablation, not a matched training condition. FLYNN (repro.) has no LiDAR input.

The lateral effect is model-dependent. With LiDAR on, FLYNN-LiDAR scores 73.4\% left-off but 91.6\% right-off (difference $-18.2$ points, $p=7.8\times10^{-14}$). MobileNet reverses this ordering: 85.4\% versus 62.6\% (difference $+22.8$ points, $p=1.1\times10^{-18}$). FLYNN (repro.) is nearly symmetric at 83.4\% and 81.0\% ($p=0.31$). The acquired graph contains slightly fewer left than right visual-column neurons (2{,}208 versus 2{,}310), but a simple left-pathway deficit would predict worse, not better, performance when the right eye is off and only left vision remains. Learned lateral dependence, camera overlap, and training-data effects remain candidate explanations; the present runs cannot distinguish them.

Collisions provide a safety measure when read with success. At full vision with LiDAR, FLYNN-LiDAR averages 2.3 collisions per episode at 96.6\% success, versus 7.7 for MobileNet, 8.3 for EfficientNet, and 5.0 for SmallWorld. Thus it has 3.3$\times$ and 3.6$\times$ fewer collisions than the CNNs at higher success. Under total loss, its scan reduces collisions from 80.5 to 40.1 while increasing success.

\subsection{Out of distribution textures}

\begin{figure}[!t]
\centering
\includegraphics[width=\columnwidth]{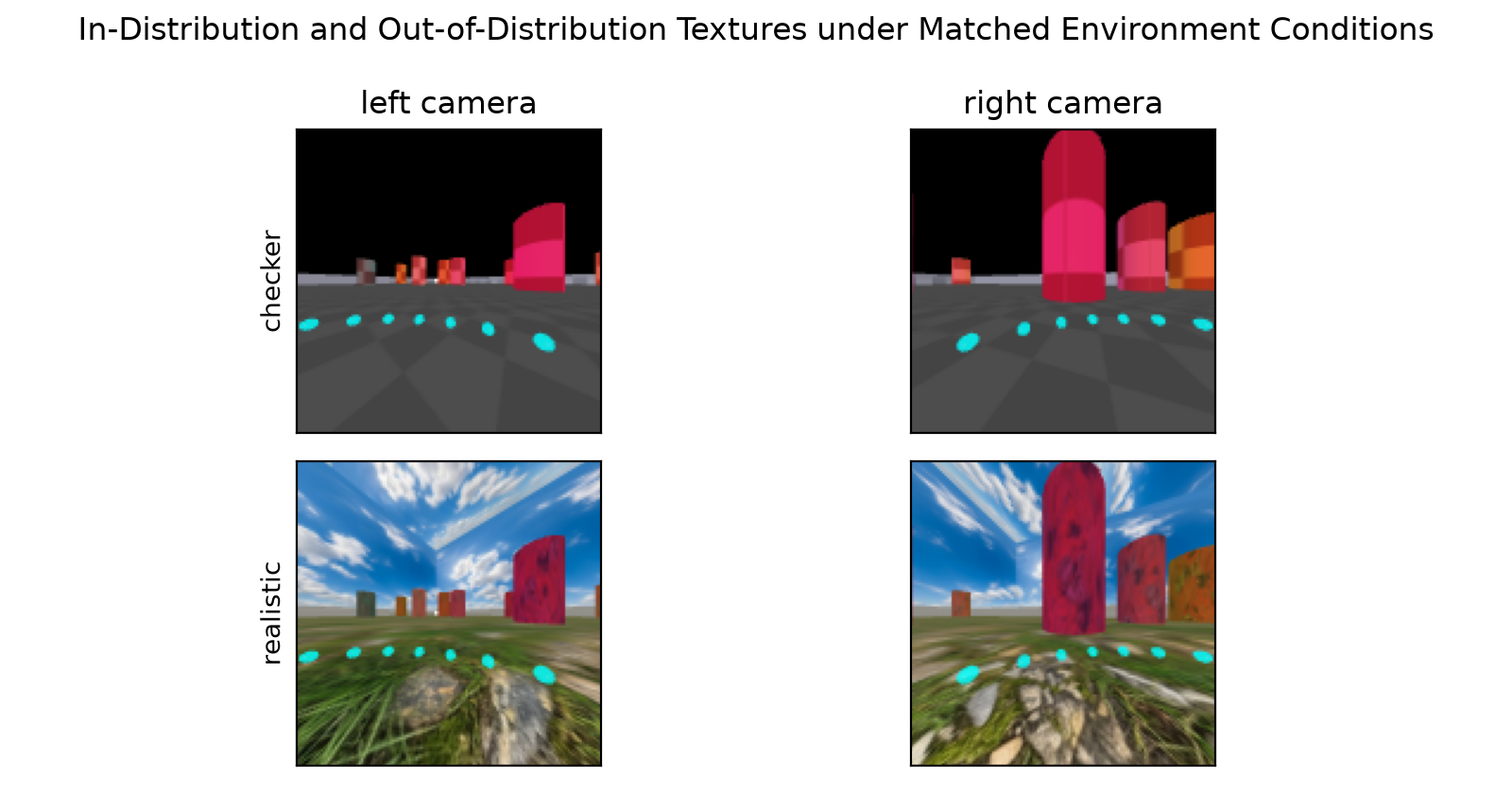}
\caption{OOD arena with photorealistic textures and paired camera frames. Cyan dots are ground hits from the 32-ray LiDAR scan (whisker strip).}
\label{fig:ood}
\end{figure}

On OOD textures (Fig.~\ref{fig:ood}), both CNNs score 0\% with or without LiDAR, and FLYNN (repro.) remains near 29\%. Among LiDAR-capable models, only FLYNN-LiDAR improves with the scan, from 44.2\% to 58.4\%. All eight comparisons against the four other models across the two OOD LiDAR states are significant at $p<10^{-6}$. SmallWorld instead declines from 26.8\% to 7.4\% ($-19.4$ points, $p=3.7\times10^{-20}$). The scan therefore does not guarantee robustness under shift.

\subsection{Blackout onset}

\begin{figure}[!t]
\centering
\includegraphics[width=\columnwidth]{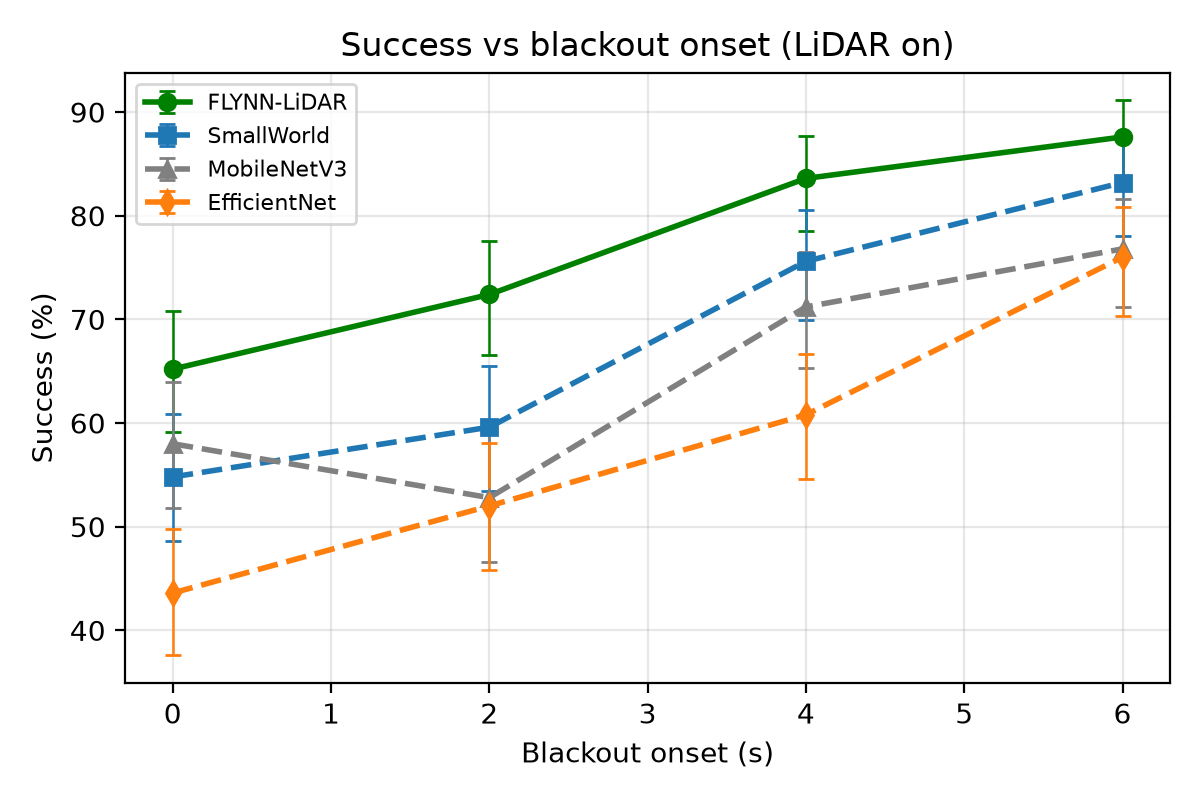}
\caption{Success versus blackout onset with LiDAR on, 250 seeds per point. Error bars are Wilson 95\% intervals. FLYNN-LiDAR is numerically highest at every onset, but not every lead is significant.}
\label{fig:blackout}
\end{figure}

FLYNN-LiDAR achieves 65.2\%, 72.4\%, 83.6\%, and 87.6\% success at 0, 2, 4, and 6 s (Fig.~\ref{fig:blackout}). Against the strongest baseline at each onset, its nominal paired-test advantage is significant at 2 s ($p=3.77\times10^{-4}$) and 4 s ($p=0.00972$), but not at 0 s ($p=0.0605$) or 6 s ($p=0.118$). For those strongest-baseline comparisons, the 2 s result remains significant after Bonferroni correction across all 12 tests (adjusted $p=0.00452$), but the 4 s result does not (adjusted $p=0.117$). MobileNet dips from 58.0\% at 0 s to 52.8\% at 2 s; the other three models improve monotonically, so dip-free recovery is not distinctive to the connectome model.

\subsection{Periodic vision flicker}

\begin{figure*}[!t]
\centering
\includegraphics[width=\textwidth]{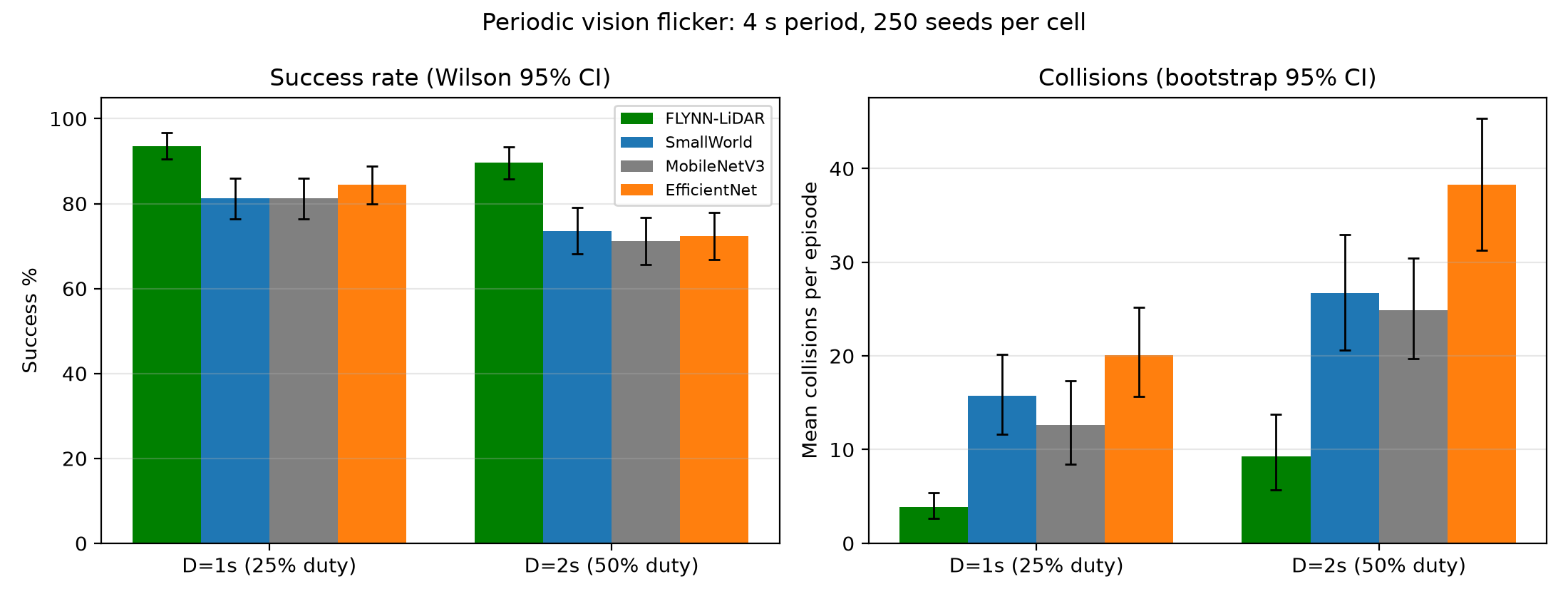}
\caption{Success and collisions under periodic vision dropout (4 s period), 250 seeds per cell. Panels show Wilson 95\% intervals for success and bootstrap 95\% intervals for collisions; the horizontal axis runs from 25\% to 50\% duty.}
\label{fig:flicker}
\end{figure*}

At 25\% and 50\% flicker duty, FLYNN-LiDAR achieves 93.6\% and 89.6\% success, respectively (Fig.~\ref{fig:flicker}). These rates are 3.0 and 7.0 points below its full-vision rate. The best baseline rates are 84.4\% and 73.6\%. Mean collisions are 3.9 and 9.3 per episode, compared with baseline ranges of 12.6 to 20.1 and 24.8 to 38.2. All six paired baseline tests are significant (nominal $p=1.1\times10^{-7}$ to $1.7\times10^{-3}$; largest six-test Bonferroni-adjusted $p=0.0100$).

\subsection{Energy}

\begin{table}[!t]
\caption{Network MACs per step and estimated energy per episode for LiDAR-on cells. CNN profiling uses one $30\times60$ grayscale backbone; connectome profiling uses the coalesced sparse graph plus readout and wind MLPs. Energy uses 4.6 pJ per MAC at 45 nm~\cite{horowitz2014energy}. Episode values are means over 500 episodes in each cell, not means across cells.}
\label{tab:energy}
\centering
\footnotesize
\begin{tabular}{lccc}
\toprule
Model & MACs/step & mJ/ep (full) & mJ/ep (total loss) \\
\midrule
\textbf{FLYNN-LiDAR} & \textbf{3.93M} & \textbf{5.9} & \textbf{6.9} \\
EfficientNet & 18.10M & 30.9 & 19.7 \\
MobileNet & 11.58M & 18.5 & 23.5 \\
SmallWorld & 5.61M & 9.1 & 8.3 \\
\bottomrule
\end{tabular}
\end{table}

\begin{figure*}[!t]
\centering
\includegraphics[width=\textwidth]{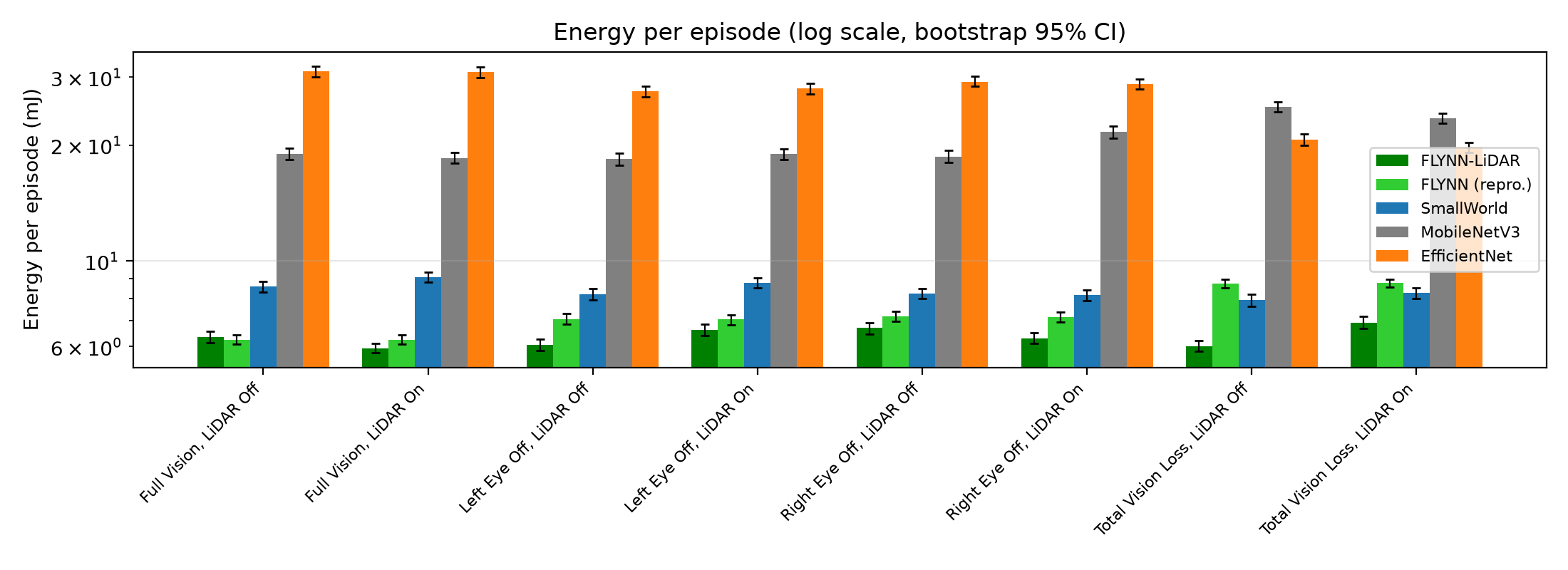}
\caption{Estimated network energy per episode from the evaluated forward-path MAC counts at 4.6 pJ per MAC (45 nm), with bootstrap 95\% intervals. Memory access, sensor acquisition, and sensor power are excluded.}
\label{fig:energy}
\end{figure*}

FLYNN-LiDAR executes 3.93M MACs per control step (Table~\ref{tab:energy}, Fig.~\ref{fig:energy}). EfficientNet and MobileNet require 18.10M and 11.58M, or 4.60$\times$ and 2.94$\times$ as many. At 4.6 pJ per MAC, the full-vision episode estimates are 5.94 mJ for FLYNN-LiDAR, 30.91 mJ for EfficientNet, and 18.52 mJ for MobileNet, giving ratios of 5.20$\times$ and 3.12$\times$. Under total loss, the estimates are 6.92, 19.70, and 23.47 mJ, giving 2.85$\times$ and 3.39$\times$. These are network-operation estimates, not measured hardware energy. Memory access, sensor acquisition, and sensor power are excluded; sparse and dense memory traffic could change the ratios in either direction, so the MAC ratios are not bounds on total energy.

\subsection{State geometry}
\label{sec:geometry}

\begin{figure*}[!t]
\centering
\includegraphics[width=\textwidth]{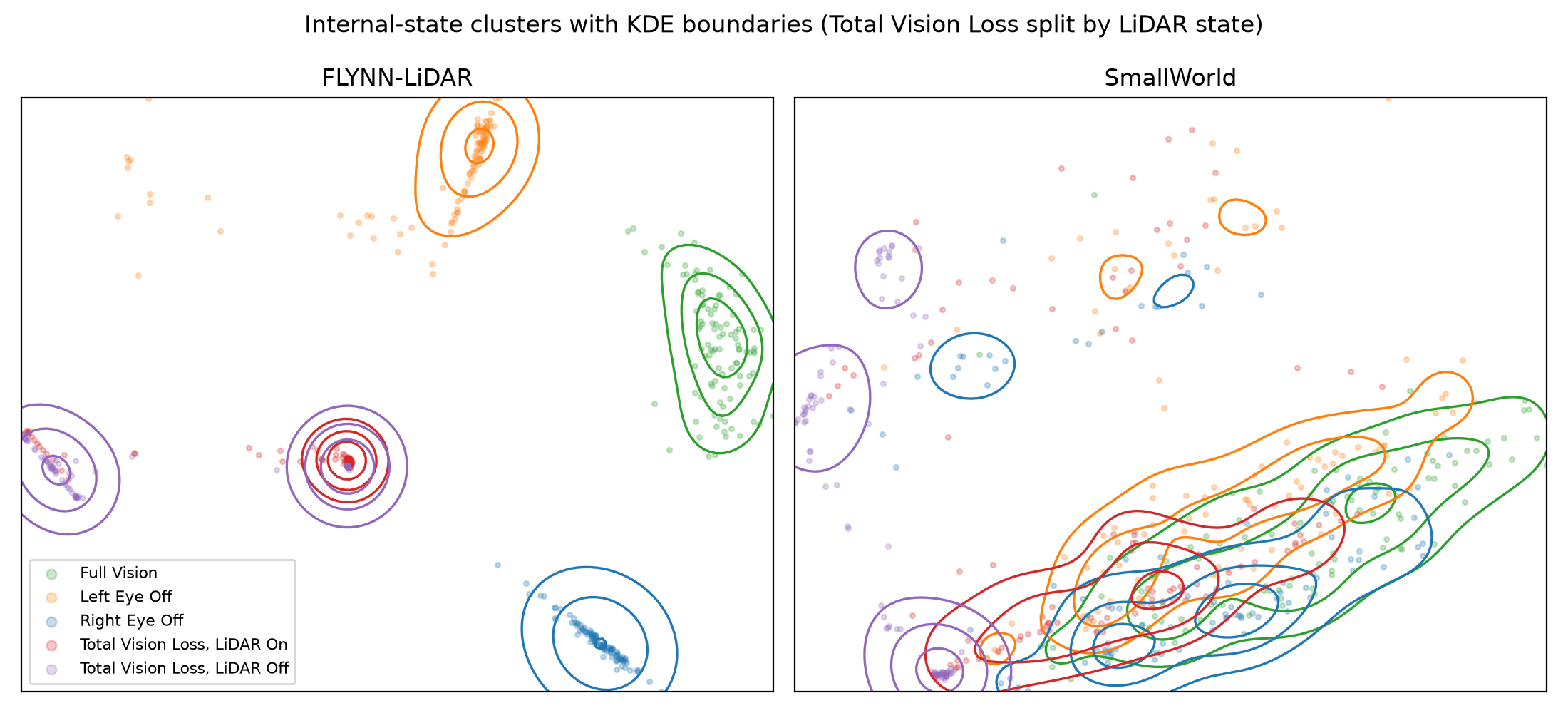}
\caption{PCA projections and KDE contours of final internal states for FLYNN-LiDAR and the SmallWorld control, 100 seeds per group. Five groups are shown: full, left-off, and right-off vision with LiDAR on; total loss with LiDAR on (red); and total loss with LiDAR off (purple).}
\label{fig:kde}
\end{figure*}

\begin{figure}[!t]
\centering
\includegraphics[width=\columnwidth]{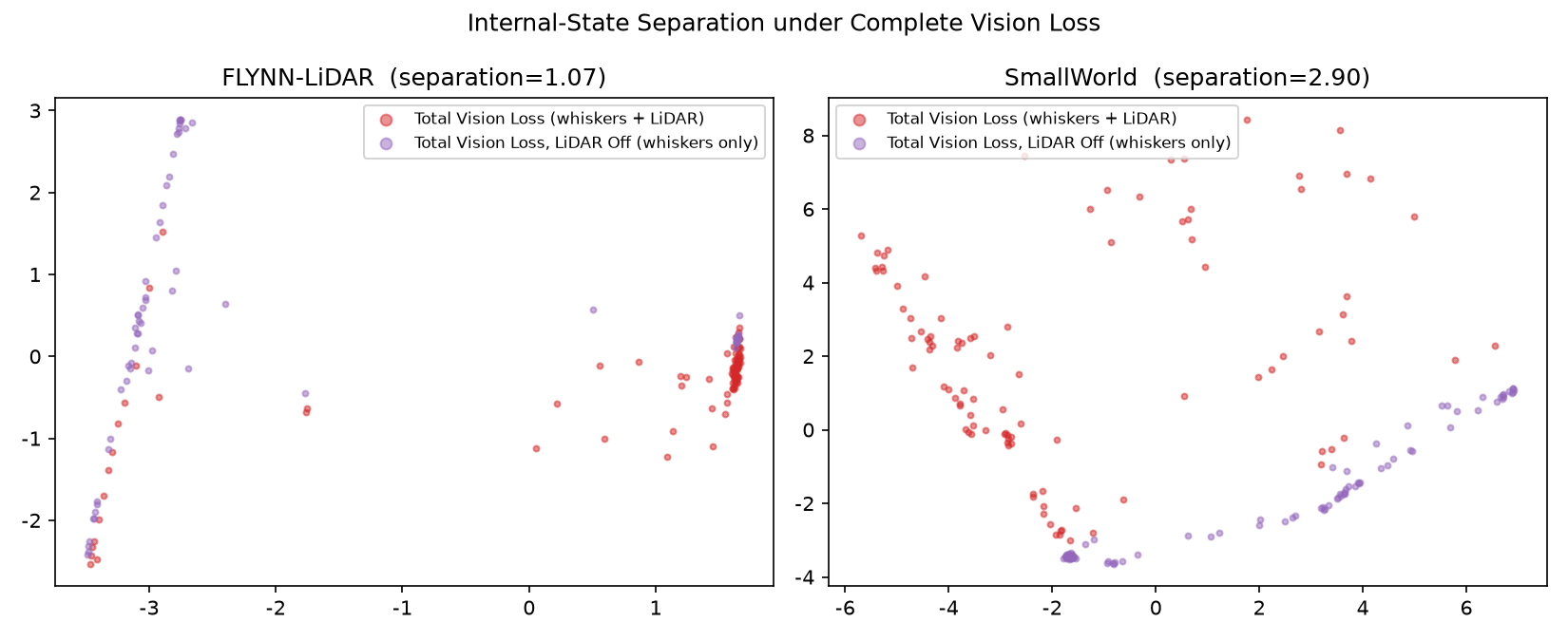}
\caption{PCA of final states under total vision loss with LiDAR on (red) and off (purple), 100 seeds per group. Full-dimensional Fisher-style scores are 1.07 for FLYNN-LiDAR and 2.90 for the SmallWorld control.}
\label{fig:pca}
\end{figure}

We first test whether the full-vision displacement is aligned with the sum of the one-eye displacements. The cosine is 0.9955 for FLYNN-LiDAR with LiDAR and 0.9859 without, 0.9616 for FLYNN (repro.), and 0.9697 and 0.9375 for SmallWorld with and without LiDAR. Thus the near-linear response persists with the scan.

Under total vision loss, full-dimensional Fisher-style mean-difference scores are 1.07 for FLYNN-LiDAR and 2.90 for SmallWorld; Figs.~\ref{fig:kde} and~\ref{fig:pca} provide two-dimensional visualizations. SmallWorld's larger displacement accompanies LiDAR gains of 17.4 points at full vision and 16.8 points under total loss in distribution, but no robustness benefit under OOD shift. These state analyses are correlational and do not establish gating, modality reweighting, or a causal mechanism.

\section{Limitations and Reproducibility}

\subsection{Limitations}

All results are in simulation with one planar robot, arena geometry, and 12 s episode cap. We evaluate one connectome instance and one training run per model, so training-seed variance and the origin of the lateral asymmetry remain unresolved. The LiDAR-off condition for FLYNN-LiDAR is an ablation of a policy trained with the scan, not a separately trained baseline. The 50 Hz scan is instantaneous and noise-free. Datasheet-parameterized range noise and 10 Hz rolling-refresh ablations are left for future work. Energy estimates omit memory and sensor costs, and state-geometry analyses are correlational.

\subsection{Reproducibility}
\label{sec:repro}

FLYNN (repro.) is our retrained reproduction, not the released FLYNN checkpoint. Its total-loss and OOD success rates are 53.4\% and 29.2\%, versus published values of approximately 44.3\% and 42.1\%~\cite{flynn2026}; its superposition cosine is 0.9616 versus 0.9998. Training budgets, evaluation seeds and protocol, and checkpoint provenance differ. It is therefore a local reference, not an exact reproduction of the published results.

\begin{table}[!t]
\caption{Realized DAgger budgets. Iteration 1 used the same recorded 500-episode teacher pool for every model instead of a new rollout; only optimizer steps are listed for that iteration. Later entries are new rollout episodes / optimizer steps.}
\label{tab:budgets}
\centering
\scriptsize
\setlength{\tabcolsep}{3pt}
\begin{tabular}{lcl}
\toprule
Model & I1 steps & Later episodes / steps \\
\midrule
FLYNN-LiDAR & 100 & 100/100 (I2-I4) \\
FLYNN (repro.) & 110 & 117/110 (I2-I3); 464/288 (I4) \\
EfficientNet & 186 & 135/186 (I2-I4) \\
MobileNet & 246 & 208/246 (I2-I3); 467/296 (I4) \\
SmallWorld & 154 & 156/154 (I2); 304/238 (I3-I4) \\
\bottomrule
\end{tabular}
\end{table}

Table~\ref{tab:budgets} records the realized budgets; no model used the nominal 500-episode, 300-step schedule throughout. The recorded pool contains 500 episodes and 176{,}394 steps, five behavior categories, and 50-step noise bursts every 60 steps.

The project repository and a browser demo are publicly available at \url{https://github.com/Ali-Mahzarnia/flynn-lidar} and \url{https://ali-mahzarnia.github.io/flynn-lidar/}. The repository includes a script that rebuilds a format-compatible graph from the public Zenodo table~\cite{dorkenwald2024zenodo}; it uses a different synapse detector from the evaluated Princeton/Codex graph and is not an exact reconstruction of that graph.

\section{Conclusion}

A fixed 32-ray projection can add range input to the fly whisker pathway with only a learned scalar input gain. FLYNN-LiDAR shows its strongest robustness under total vision loss, OOD textures, and periodic flicker, while MobileNet retains a left-eye-specific advantage and the lateral effect remains unexplained. Profiling the evaluated paths yields 4.60$\times$ and 2.94$\times$ per-step MAC ratios against the CNNs, with lower estimated episode energy but not an orders-of-magnitude advantage. Multi-seed training, realistic scan noise and refresh, and hardware evaluation are necessary next steps.

\bibliographystyle{IEEEtran}
\bibliography{refs}

\end{document}